\documentstyle[12pt]{article}
\title{Mass-radius relations for helium white dwarfs
\thanks{No.of copies:3, No.of pages:7, No. of figures:1, No. of tables:0}}
\author{R. Civelek, N. K{\i}z{\i}lo\u{g}lu
             \thanks{Proof and
            offprints to: R. Civelek, Physics Dept.
            Middle East Tech. Univ. 06531 Ankara, Turkey}\\
                                                                \\
Physics Department\\
Middle East Technical University\\
Ankara 06531, Turkey}
\date{June 25, 1998} 
\begin{document}
\begin{titlepage}
\maketitle
\end{titlepage}

\begin{abstract}
We studied the static mass-radius relation of white dwarf stars 
with masses greater than 0.45 $M_\odot$. We assumed pure
degenerate helium interiors at a finite temperature with luminosity-mass
ratio due to neutrino energy loss. We compared the obtained M-R relation with
those of other writers. 

           
\end{abstract}
\newpage
\section{Introduction}
In recent years,  white dwarf stars are receiving increasing
attention.
The theoretical relationship between the mass and radius of a white
dwarf is important in interpreting some of the observational results.
M-R relation was first defined by Chandrasekhar (1939). Later Hamada and
Salpeter (1961) obtained numerical models for different core
compositions by considering a fully degenerate configuration
(zero-temperature) because of their higher densities ($10^{6}$-$10^{8}$
g $cm^{-3}$).

A real WD is not a zero-temperature object. The inclusion of hydrogen
envelope increases the radius depending on the amount of hydrogen
present which is not known with certainty (Koester, 1987; Hamada and
Salpeter, 1961). Hamada and Salpeter (1961) mentioned that for massive
WDs ($M_\odot$ $>$ 0.7) non degenerate envelope is rather insignificant.
Benvenuto and Althaus (1998) in their recent studies also concluded that
thick H envelopes increase the radii especially in the case of low mass
WDs.

The recent studies on the relation of M-R for the WDs are those of Wood
(1990), Vennes et al. (1995) and Althaus and Benvenuto (1997, 1998).
Vennes et al. (1995) computed static M-R relation for masses between 0.4
$M_\odot$ and 0.7 $M_\odot$ assuming non zero temperature effects. They
assumed the luminosity is proportional to the mass which works for cool
WDs but their results are in the range of high effective temperatures.
As Althaus and Benvenuto (1997) mentioned, luminosity is not
proportional to a constant for hot WD interiors because of neutrino
emission. One must include neutrino cooling which causes larger radii
for WDs. Neutrino losses are important especially for masses greater
than 0.4 $M_\odot$ (see Fig.11 of Althaus and Benvenuto, 1997).

The purpose of this study is to present the effect of neutrino emission
at finite temperatures.
We considered fully degenerate configuration for
WDs with pure helium composition to obtain M-R relation for masses
greater than 0.4 $M_\odot$ with neutrino emission taken into account as
well. In section 2 we describe the procedure that we follow. 
 In section
3 we present and discuss the results and compare the obtained M-R
relation with the other results.

\section{Procedure}
Our stellar WD models are calculated on the assumption that the WD is
spherically symmetric and in hydrostatic equilibrium. Then, four stellar
structure equations that must be satisfied by this structure are
integrated outward with Runga-Kutta iteration tecnique. For the equation
of state, we followed the procedure given by Althaus and Benvenuto
(1997) for a dense plasma in which the electrons are strongly degenerate
at a finite temperature, that is we included in the equation of state
Coulomb interaction, 
Thomas-Fermi deviation from uniform charge distribution of the electrons
and the exchange contribution to the free energy at finite temperature.
As far as neutrino losses are concerned we considered  photo
neutrino process (Itoh et al., 1989), plasma neutrino process (Itoh et
al., 1989; Itoh et al., 1992)
 and neutrino
Bremstrahlung for the liquid phase (Itoh and Kohyama, 1983).
For the conductive opacities we used the analytic fits given by Itoh et
al. (1983) for high densities.

\section{Results}
In this  paper, we give the first results of our white dwarf models
including neutrino emission, that is M-R  relations of the
fully degenerate helium WDs for masses greater than 0.4 $M_\odot$ are
presented.

Figure 1. shows the M-R relations for helium WDs calculated
 by Hamada and Salpeter
(1961), Althaus and Benvenuto (1997), Vennes et al. (1995) and our
results. 
In this figure, plus sign shows our results obtained at the temperature
$T_{c}$= $10^{7}$ K using the density values given in Table 1A of Hamada
and Salpeter (1961) for helium core (cross sign shows the results of
Hamada and Salpeter). The difference in mass and radius
is about 1\%. As seen from the figure the static mass-radius relations
are
modified by thermal effects and neutrino emission particularly for WDs
of low mass. These thermal
effects and neutrino emission can cause deviations from the
zero temperature M-R curve which are almost of the same order as
Hamada-Salpeter corrections to the standard Chandrasekhar M-R curve.
 We repeated the calculations for a
different central temperature $T_{c}$= 5x$10^{7}$ K. The resulting data
is also shown (circle sign) in the figure.
     The increase in $T_{c}$ causes the curve to
shift upward for  the masses smaller than 0.7 $M_\odot$. We plot on the
same figure also the results of Althaus and Benvenuto (1997) for zero
temperature (star sign) and for $T_{c}$= 5x$10^{7}$ K helium WD models
(full box sign). The models of Vennes et al. (1995) for masses smaller
than 0.7 $M_\odot$
are shown by open box sign. Their models have larger radii than the
other models plotted in the same figure due to the high effective
temperature (49000 K) they used which means higher central temperature.

In our calculations with internal temperature of 5x$10^{7}$ K,
 we found R=0.015 $R_\odot$ for 0.5 $M_\odot$ WD. The
effective temperature of 0.5 $M_\odot$ WD star can be
assigned to be in the range 17x$10^{3}$ - 17.5x$10^{3}$ K using the results of
Atweh and Eryurt-Ezer (1991) for the lower boundary of convection zone
of helium WDs in the case of strong convection. They give the depth of
convection zone for this effective temperature range between 30 and 60
km. Therefore, about 7\% difference in the radii of 0.5 $M_\odot$ WDs at
the mentioned interior temperature, between our study and the study of
Althaus and Benvenuto, is not due only to the absence of helium
atmosphere in our study but also due to neglecting star's thermal
history. Detailed evolutionary models are necessary for better
interpretation of observations of WDs.

Acknowledgement. We would like to thank Prof. Dr. D. Eryurt-Ezer for her
suggestions in this study.

\newpage
\section*{Figure captions}

Fig.1. Mass-radius diagram for pure helium white dwarfs.
Plus and empty circles are the results of the present study for central
temperatures of $10^{7}$ K and 5x$10^{7}$ K,
 respectively. Star and full box signs show the results of Althaus and
Benvenuto (1997) for zero-temperature and $T_{c}$=  5x$10^{7}$ K WD
stars, respectively. Also shown are the data of Hamada and Salpeter
(1961) (cross sign) and Vennes et al. (1995) (empty box sign). 

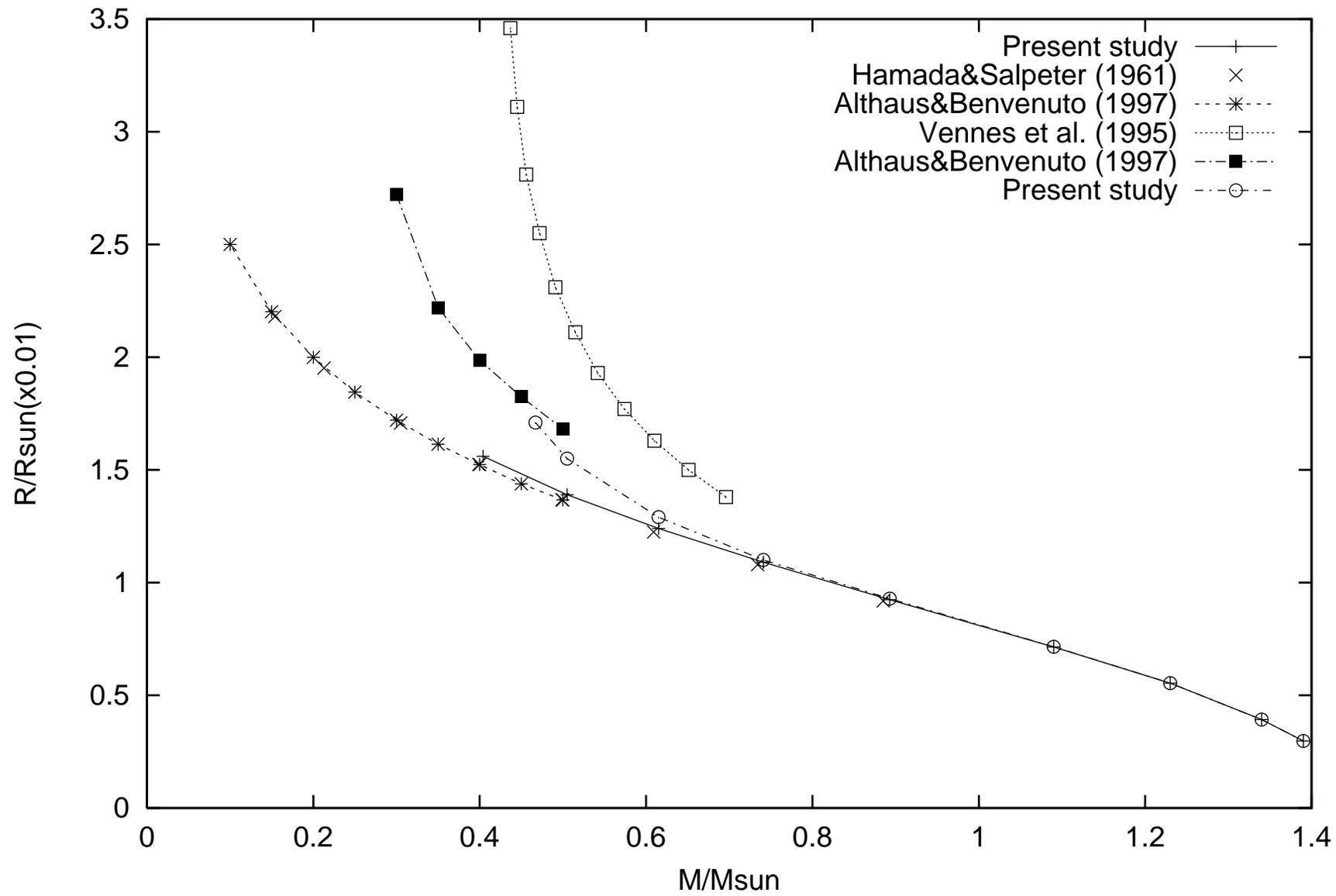
\end{document}